\documentclass[11pt]{article}

\newcommand{\kim}{ k_{1\mu}}                                      
\newcommand{\kom}{ k_{0\mu}}                                      
\newcommand{\ki}{ k_{1}}
\newcommand{\yn}{ Y_{n}}

\newcommand{\kt}{ k_{2}}                                             
\newcommand{\ko}{ k_{0}}

\newcommand{\kin}{ k_{1\nu}}                                      
\newcommand{\kor}{ k_{0\rho}}                                      
\newcommand{\kon}{ k_{0\nu}}                                      
\newcommand{\ktm}{ k_{2\mu}} 
\newcommand{\ktn}{ k_{2\nu}}

\newcommand{\ddS}{\frac{\delta}{\delta \Sigma}}

\newcommand{\eps}{ \epsilon}

\newcommand{\kib}{\mbox {$ \bar{k_{1}}$}} 
\newcommand{\ktb}{\mbox {$ \bar{k_{2}}$}}

\newcommand{\be}{\begin{equation}}
\newcommand{\br}{\begin{eqnarray}}
\newcommand{\ee}{\end{equation}} 
\newcommand{\er}{\end{eqnarray}}

\begin{document}
\renewcommand{\theequation}{\thesection.\arabic{equation}}

\title{
\hfill\parbox{4cm}{\normalsize IMSC/2005/03/02\\
                               hep-th/0503011}\\        
\vspace{2cm}
Loop Variables and the Interacting Open String in a Curved Background.
\author{B. Sathiapalan\\ {\em Institute of Mathematical Sciences}\\
{\em Taramani}\\{\em Chennai, India 600113}\\ bala@imsc.res.in}}           
\maketitle     

\begin{abstract} 
Applying the loop variable proposal to a sigma model (with boundary)
 in a curved target
 space, we give
a systematic method for writing the gauge 
and generally covariant interacting equations of motion for the modes of the
 open string in a curved background. As in the free case described in an earlier paper,
 the equations are obtained
by covariantizing the flat space (gauge invariant) interacting equations and then 
demanding gauge invariance in the curved background.  The resulting equation has the form
of  a sum of terms that would individually be gauge invariant in flat space or at zero
interaction strength, but mix amongst themselves in curved space when interactions 
are turned on. The new feature is that the loop variables are deformed so that there 
is a mixing of modes. Unlike the free case, the equations are coupled, and all the modes of the
 open string are required for gauge invariance.

\end{abstract}

\newpage

\section{Introduction}

In an earlier paper \cite{BSCB} (hereafter ``I'') the loop variable
proposal \cite{BSLV} \footnote{A review of the loop variable approach is given in
\cite{BSREV,BSGI}.} was extended to free open strings in curved space. A systematic
method of writing down gauge and generally covariant 
equations for the massive modes of the open string was given. The basic method
was to take the loop variable equation in flat space and apply it
to curved space by making the loop variables conjugate to
 Riemann normal coordinates. This then gives a covariantized version
of the flat space equation. The problem then is that if one tries to map
the gauge transformations of the loop variables to gauge transformations of 
space time fields in the 
usual manner, one finds that the space time fields do not have well defined 
gauge transformations.  In the loop variable approach this is a central
issue. The equations written in terms of loop variable momenta are always
gauge invariant. What is non trivial is the existence of a well defined
map from loop variables to space time fields in such a way that gauge 
transformations are also well defined on the space time fields.
If this can be achieved then we have gauge invariant equations in space time.

Thus, if we let $L$ be the loop variable expression and $S$ be 
the corresponding expression in terms of space time fields,
\[
{\cal M} : L \rightarrow S 
\]
\be    \label{1}
{\cal M} : L^g (= L+ \delta L)\rightarrow S^g= S+\Delta S 
\ee
where 
\be 
{\cal G}: L \rightarrow L^g
\ee
 is the gauge transformation of $L$. These two equations define
the action $\cal G$, gauge transformation on the space time fields: the gauge transformation
should be such that 
\be   \label{1.1}
{\cal G} : S \rightarrow  S+\delta S = S^g = S+ \Delta S
\ee
where $S^g$ is defined by (\ref{1}) as the map of $L^g$.
In curved space space the map $\cal M$ is the one obtained
by covariantizing the flat space map using Riemann normal coordinate
method mentioned above. General Covariance demands that 
the gauge transformation 
of space time fields is the covariantized version
of the flat space transformation with possible
additions of curvature dependent terms. 
In the free case as discussed in I, it is easy to see
that given these two constraints it is not possible to satisfy (\ref{1.1}). 
 The solution was to modify the map from loop variables
to space time fields in such a way that with the same 
gauge transformations (\ref{1.1})is satisfied.
 Thus we change $\cal M$ to $\cal M'$:
\[
{\cal M'} : L \rightarrow S'= S+ S_1
\]
\be
{\cal M'} : L^g (= L+ \delta L)\rightarrow S^{'g}=S + S_1 + \Delta S 
\ee 
Notice that the map on $\delta L$ is unchanged. 
Now we have
\be   
{\cal G}:S'(=S+S_1)\rightarrow S+S_1 + \delta (S+S_1)
\ee
and we want 
\be  \label{2}
\delta (S+S_1) = \Delta S
\ee
 What was shown in \cite{BSCB}
was that it is always possible to find $S_1$ such that (\ref{2})
is satisfied.  The gauge transformations are always just the usual 
 gauge transformations, covariantized for curved space.
As an example of this, take the case where 
\[
L= k_{0\rho} \kim \kin
\]
\[
S=D_\rho S_{1,1 \mu \nu}
\]
Then 
\[
\delta L = k_{0\rho} \lambda _1 k_{0(\mu}k_{1\nu )}
\]
Thus
\be   \label{1.3}
{\cal M} : \delta L \rightarrow \Delta S =  D_\rho D_{(\mu }\Lambda _{1,1\nu )}+ 
{1\over 3}(R^\alpha _{~\nu \rho \mu} + R^\alpha_{\mu \rho \nu})
\Lambda_{1,1\alpha}
\ee
On the other hand 
\be       \label{1.4}
\delta S = D_\rho (\delta S_{1,1\mu \nu})=D_\rho D_{(\mu}\Lambda _{1,1\nu )}
\ee
We see a mismatch between (\ref{1.3}) and (\ref{1.4}).

Following \cite{BSCB} we let 
\[
S_1 ={1\over 3}(R^\alpha _{~\nu \rho \mu} +
 R^\alpha_{\mu \rho \nu})( S_{2\mu} - {D_\mu S_2 \over 2k_0^V})
\]
and using   
$\delta (  S_{2\mu} - {D_\mu S_2 \over 2k_0^V})=\Lambda _{1,1\mu}$ 
\footnote{The gauge transformations of fields are taken from I}
we see that (\ref{2}) is satisfied.
Thus we have a well defined map from loop variables to space time fields.
When this is done the space time equations 
are automatically gauge invariant - because they are obtained by a
well defined map from an expression that is gauge invariant in terms
of loop variables. 

The above was an outline of the discussion of
free open string
fields in curved space, that was worked out in I. 

In this paper we extend this to the interacting case. 
In the curved space interacting case, there is a further complication, 
 that makes the flat space technique used for interactions
inadequate. This is essentially the same problem that one encounters
in the free case, i.e. that of covariant derivatives in curved space 
making it difficult to define gauge transformations. Thus one has to
generalize the technique that was used to define gauge transformations 
in the flat space interacting case. The technique there was to recursively 
define the gauge 
transformation of the higher modes so that the offending terms (i.e.
terms involving lower modes 
that cannot
be attributed to gauge transformations of the lower modes), are
absorbed into the gauge transformation laws of higher modes, 
thereby making the map from
loop variables to space time  always well defined. 
In the notation used above, the gauge transformation of the highest
mode in $S$ is defined so that $\delta S = \Delta S$.
 It was shown in
\cite{BSGI,BSREV} that there is a systematic recursive
way of doing this. We will use 
the same technique here. However it turns out that if one attempts to do this
exactly as in the flat space case, often there {\em is} no higher mode that
 can absorb the offending terms! This is solved by deforming the loop variables
so that higher modes always exist. Once this is done, the same procedure
that was used in flat space works here also. This deformation is very similar
to the deformation that takes us from ordinary derivatives to covariant
derivatives. In our case all the loop variables (which are really 
generalized momenta)  have to be deformed. 

The above was an outline of the procedure.  One sees that the main idea,
which is the same in the free case, is to use general covariance and gauge
invariance to obtain the equations, starting from the flat space equations.
If this procedure is unique then the answer one obtains should be the correct
one for string theory. This can be verified by doing explicit curved space 
sigma model RG calculations. 
This is an important issue but is not addressed        
in this paper.
 
This paper is organized as follows. Section 2 contains a short review 
of the interacting flat space case \cite{BSGI}.
Section 3 describes the problems one encounters when attempts
to go to curved space and the resolution of these problems. An outline
of a sample calculation is given. The full details have not been worked out
because they are very tedious and not particularly illuminating. Section 4 
contains
a summary and some conclusions. We have not included a review of the loop variable
approach. This paper is a follow up of I and should ideally be read in conjunction
with I.

\section{Review of Interacting Case (Flat Space)}

The flat space interacting case was described in \cite{BSGI,BSREV}.
There are two main steps. The first is obtaining the equations for the loop
variable momenta, and the second is mapping this to an equation for
space time fields and then defining the gauge transformation laws
for the space time fields under which these equations are invariant.

{\bf Step 1:}  The equations are obtained by first Taylor expanding all
the vertex operators about one point on the world sheet (say z=0), so that
the loop variable looks like that of the free theory. Note that
this is legitimate only because of the presence of a world sheet
cutoff. The cutoff is kept finite except when we go to the on-shell
limit. All the complications
 are hidden in the $z$-dependent generalized momenta $k_n(t,z(t))=k_n(t)+
a_1zk_{n-1}(t) + a_2z^2 k_{n-2}(t)+....$. We can simplify things by
letting $z(t)=t=z$ and using $z$ to label the momenta as well as
label the position on the world sheet. Thus $k_n(t,z(t))$ can be
denoted by $\bar k_n(z)$. Then one
applies the same technique as for the free case, which is to perform
the operation $\ddS$ on the loop variable and set it to zero. 
An example of a free loop variable equation is: 
\be
(-2\ki .\ko \kt .\ko + \ki .\ko \ki .\ki +2 \kt .\ki \ko .\ko )i\kom =0 
\ee
There is an overall factor of $(\eps ^2)^{\ko ^2-1}$ multiplying it.
If we replace $k_n$ by $\bar k_n(z_i)$ we get the interacting loop equation. 
The first term in this equation after performing the dimensional reduction becomes: 

\[
(\eps ^2)^{\ko ^2}(-{4\over \eps ^2})[\kib (z_1) .\ko (z_2) \ktb (z_3) .\ko (z_4) 
+ 
\]
\be  \label{1.41}
2 \bar k_3 (z_1).\ko (z_2)
k_0^V (z_3)k_0^V(z_4) + k_3^V(z_1)k_0^V(z_2)k_0^V(z_3)k_0^V(z_4) ]i\kom (z) =0
\ee
Integrations over all the $z_i$ are understood.
Here $\eps$ is the world sheet cutoff. 

The equations 
obtained by this procedure are gauge invariant under 
\be   \label{2.5}
\bar k_n(z)\rightarrow \bar k_n(z)+\int dz' ~\lambda _p(z')\bar k_{n-p}(z)
\ee
provided we use the tracelessness constraint on the gauge parameters ($\lambda _p k_n .k_m =0$ when both
$n,m \neq 0$).

{\bf Step 2:}  The loop variable expressions are mapped to space time fields
using relations of the form:
\be    \label{3}
<k_{m\mu}(t_1)k_{n\nu}(t_2)> = S_{m,n\mu \nu}\delta (t_1-t_2)+ S_{m\mu}
S_{n,\nu}
\ee
and it's obvious generaliztions.
Using these one can work out 
\[
\int dz_1~dz_2~
<\bar k_{m\mu}(z_1)\bar k_{n\nu}(z_2)> 
\]
\[ 
=\int dz_1~dz_2~
<(k_{m\mu}(z_1)+ a_1z_1k_{m-1\mu}(z_1)+...)(k_{n\nu}(z_2)+a_1k_{n-2\nu}(z_2)+
...)>
\]
\be    \label{4}
= \int dz_1~dz_2~
[S_{m,n\mu \nu}\delta (z_1-z_2)+ S_{m\mu}S_{n,\nu}
+ a_1z_1S_{m-1,n\mu \nu}\delta (z_1-z_2)+....]
\ee

We can now work out the gauge transformations. Take (\ref{4}) as an example.
The gauge transformation of the LHS can be worked out using (\ref{2.5}). The 
gauge transformation of the fields on the RHS are chosen to satisfy this 
equation. This is done recursively: The gauge transformation of the highest
level field ($S_{m,n\mu \nu}$) is fixed by this equation in terms of the
gauge transformations of the lower level field. This gives a systematic way 
of obtaining the gauge transformation laws of all the fields.

\section{Interactions in Curved Space}

The above method needs to be modified in curved space for the reason mentioned
in the introduction. Consider the term 
\be   \label{5}
\kor k_{m\mu}k_{n\nu}
\ee
When we map to space time fields, the derivative becomes a covariant
derivative. Furthermore as in the free case, the gauge transformation of
this entire expression will not be given by the covariant derivative of the
gauge transformation acting on 
\be     \label{6}
<k_{m\mu}k_{n\nu}>
\ee
 What this means is that (\ref{5}) and (\ref{6}) have to be treated as
separate independent entities if gauge transformations are
to be consistently defined.  But this is inconsisent with the idea of
$k_0$ being a derivative. In the free case we resolved this problem by changing
the map $\cal M$. Here that is not possible - the two have to be treated
independently and we must give up the idea that $k_0$ is a 
 (generally covariant) derivative. 

One modification that achieves this without spoiling the simplicity
of the gauge transformation structure is to deform the $k_n$ to 
\be   \label{def}
p_{n\nu} = k_{n\nu} + g[k_{(n+1) \nu}k_{0V}-k_{(n+1)V}k_{0\nu}]
\ee
$g$ can be taken proportional to the coupling constant for the simple
reason that this deformation is not required when the coupling constant
is zero.
It is easy to see that $p_{n}$ has the same gauge transformation as $k_n$.
The antisymmetrization with $k_{0V}$ is required in order that $p_{n}$ be
invariant under $\lambda _{n+1}$ transformations. Under this deformation
$p_{0\mu} =\kom + g[\kim k_{0V} - k_{1V}\kom]$. This makes $p_0$
similar to a gauge covariant derivative, with $\kim$ the gauge field. 
The strategy is to use the $p_n$ 's in place of $k_n$ in the loop
variable. Thus we have $e^{i(\sum _n p_{n}\yn )}$ as the loop variable
as far as step 1 is concerned. For step 2 we use the same procedure as before
and recursively define gauge transformations for the highest level field
in each expression. Whenever we have a space time derivative it is accompanied
by a higher level field. Thus $k_n k_m$ becomes $p_{n} p_{m}$
 where the highest
level field is $S_{n+1,m+1}$. On the other hand $k_0 k_n k_m$ becomes
$p_0p_{n}p_{m}$ where the highest level field is $S_{1,n+1,m+1}$. Thus
it becomes an independent expression, not simply the covariant derivative
of the other expression.

\subsection{Example}

Consider the following term (we considered this in the free case as well):
This term is level 2 and is simpler than the terms occurring in (\ref{1.41})-
even so the space time expression that it is mapped to is very complicated.
We will therefore only outline the steps involved in constructing the map to space time
fields and the gauge transformation of the fields:
\be
L=\int \int \int dz_1dz_2dz_3~ k_{0\rho} (z_1) \bar \kim (z_2) \bar \kin (z_3)
\ee
First we replace it by the deformed version replacing $k$ by $p$:

\be   \label{7}
L=\int \int \int dz_1dz_2dz_3~ p_{0\rho} (z_1) \bar p_{1\mu} (z_2) \bar 
p_{1\nu} (z_3)
\ee

We then write 
\[
\int dz ~\bar p_{n\nu}(z) = 
\]
\be    \label{8}
\int dz ~\bar k_{n\nu}(z) + g[\int dz~
\bar k_{(n+1) \nu}(z)\int dz'~k_{0V}(z')-\int dz~\bar k_{(n+1)V}(z) \int dz'~ 
k_{0\nu}(z')]
\ee

and substitute in (\ref{7}).  This gives 27 terms and we will not
write them all out. We write down some terms to illustrate the procedure.
Integrals over all $z$'s are understood:
\[
p_{0\rho}(z_1)\bar p_{1\mu}(z_2)\bar p_{1\nu}(z_3)~=~
( k_{0\rho}(z_1) + g[\bar k_{1 \rho}(z_1)k_{0V}(z_1')-\bar k_{1V}(z_1) k_{0\rho}(z_1')] )
\]
\[
(\bar k_{1\mu}(z_2) + g[\bar k_{2 \mu}(z_2)k_{0V}(z_2')-\bar k_{2V}(z_2) k_{0\mu}(z_2')] )
\]
\be
(\bar k_{1\nu}(z_3) + g[\bar k_{2 \nu}(z_3)k_{0V}(z_3')-\bar k_{2V}(z_3) k_{0\nu}(z_3')] )
\ee

The leading term amongst the 27 is
\be    \label{9}
 k_{0\rho}(z_1)\bar k_{1\mu}(z_2)\bar k_{1\nu}(z_3)
=k_{0\rho}(z_1)(k_{1\mu}(z_2)+z_2k_{0\mu}(z_2))
(k_{1\nu}(z_3)+z_3\kon (z_3))
\ee

A term of O($g^3$) is
\[
g^3 \bar k_{1\rho}(z_1)\bar k_{2\mu}(z_2)\bar k_{2\nu}(z_3)
k_{0V}(z'_1)k_{0V}(z'_2)k_{0V}(z'_3)
\]
\[
= -1(k_{1\rho}(z_1)+z_2k_{0\rho}(z_1))
(\ktm (z_2)+z_2\kim (z_2) + {z_2^2\over 2}\kim(z_2))
\]
\be   \label{10}
(\ktn (z_3)+z_3\kin (z_3) + {z_3^2\over 2}\kin(z_3))
\ee

The value of $k_{0V}$ is always one less than 
the renormalization group dimension
of the vertex operator multiplying $p_n$ which is $n-1$\cite{BSGI,BSREV}.
Thus $k_{0V}$ adds up to -1. This gives the factor -1. 

We then map to space time fields usig $\cal M$.
\be
{\cal M}: L \rightarrow S
\ee

The $z$-independent term  of the sum (\ref{9}) maps to
\be
\int _0^a ~dz_1~\int _0^a ~dz_2~\int _0^a ~dz_3~
<\kor (z_1) \kim (z_2) \kin (z_3)> ~=~ 
D_\rho (aS_{1,1\mu \nu} + a^2A_{1\mu}A_{1\nu}) 
\ee

Similarly the $z$-independent term in (\ref{10}) maps to

\[
\int _0^a ~dz_1~\int _0^a ~dz_2~\int _0^a ~dz_3~
<k_{1\rho} (z_1) \ktm (z_2) \ktn (z_3)> = aS_{1,2,2\rho \mu \nu} +
\]
\[ 
a^2(S_{1,2\rho \mu}S_{2\nu}+S_{1,2\rho \nu}S_{2\mu}+S_{2,2\mu \nu}A_{1\rho})
\]
\be
+ a^3 A_{1\rho}S_{2\mu}S_{2\nu}
\ee
Note that the highest level field $S_{1,2,2\rho \mu \nu}$ 
in this expression has no derivative acting on it. Other fields at the same
level such as $S_{1,2,2\rho \mu V}$ have derivatives acting on them.

The above illustrates the calculation of $S$ from $L$. 

The gauge transformation
of $L$ gives $L^g$, and its map to $S^g$ can be similarly evaluated. This gives
$\Delta S$ in the notation used for flat space. 
The gauge transformation of $S$ will involve $\delta S_{1,2,2\rho \mu \nu},
\delta S_{1,2,2 \rho \mu V},~ \delta S_{1,2,2\rho VV},~ \delta S_{1,2,2V\mu \nu},~
 \delta S_{1,2,2VVV}$ in addition to $\delta S_{2,2\mu \nu},~
 \delta S_{1,2,\rho \mu},~...$ and all the lower fields.  This will give
$\delta S$. 
As before, we require that $\Delta S = \delta S$.
Thus we can use this
to define $\delta S_{1,2,2\rho \mu \nu}$ (which, as pointed out above has
no derivatives acting on it) in terms of the other fields.
This defines a recursion both in terms of the level of the fields and the number
of indices in the non-V directions. Note that $\delta S_{1,2,2VVV}$ will involve
{\em only} lower level fields, and $\delta S_{1,2,2\rho VV}$ 
will involve $\delta S_{1,2,2VVV}$
and lower fields, and so on.

 This is the same kind of recursive
scheme that was used in flat space. The difference is that the derivative
terms have been incorporated into a separate independent term in the scheme 
by means of the deformation.

Finally, instead of using $\cal M$ we can use $\cal M'$ which differs
in its action on the free part of the map as explained in the introduction.
 
To summarize, by means of the deformation  (\ref{def}), one is able to
write down a system of generally covariant equations and a corresponding 
set of gauge transformations that leave these equations invariant. These
equations have the form of the covariantized flat space equations ($S_0$
)added to
which are terms involving higher level modes, ($S_h $) .  
They can schematically be written as $S_0 + S_h =0$.
These terms are necessary
for gauge invariance in curved space, i.e. in curved space
$\delta S_0 = - \delta S_h \neq 0$.  
In flat space, the equations break
up into the usual equations plus gauge invariant pieces involving higher
level modes, i.e $\delta S_0=0=\delta S_h$. Presumably the system of equations can be
truncated consistently to the original flat space equations. Clearly, as a result
of the deformation, the equations obtained here are some linear combinations
of the equations obtained in \cite{BSGI} for the flat space case. 

To the extent that the equations for loop variables are RG equations
we have followed string theory prescriptions. The map to space
time fields is then fixed by considerations of gauge invariance and general
covariance - not the one that one would normally use in string theory
(because of the deformation). However
if it is true that the final set of equations are unique
 (given the particle content, gauge invariance and generally covariance), 
upto field redefinition, 
then this must be the answer that string theory 
also gives.

\section{Conclusions}

In this paper we have given a systematic method of writing down interacting
gauge invariant and generally covariant equations for all the modes of the
open string. The method is a combination of the methods used in \cite{BSCB}
for the free open string in curved space and \cite{BSGI} for interacting
open strings in flat space. It consists of the following ingredients:

1. Riemann Normal Coordinates are used for the sigma model in curved space
and generally covariant equations are obtained from the loop variable equation
 as described in \cite{BSCB}

2. The loop variable momenta are deformed as described in (\ref{def}), so the
loop variable equation is different from the flat space loop variable 
equations.

3. The definition of gauge transformation of all fields can be defined
 unambiguously using the same technique as used in the flat space case.
Once the loop variable equations are obtained in terms of $p_n$, we reexpress
them in terms of $k_n$ and then proceed exactly as in the flat space case
to define gauge transformations recursively for the higher level modes in
terms of lower level modes.

We have illustrated the various steps involved by examples. However,
the full answer (while reasonably simple in terms of
the deformed loop variables), is in even the simplest case,
 quite complicated
when written out in terms of space time fields. It seems unlikely that
working with explicit space time field equations will lead to any
insights or solutions. We suspect that it is better to work with 
loop variables directly. This is a problem for the future.
Perhaps the most interesting conclusion for the present,
is just the fact that
there exists an algorithm for writing
 down gauge and generally covariant interacting equations
for all the higher mass and spin modes of the open string in curved
space.

It would be very useful to show by explicit computations on the sigma model
that these equations are indeed equivalent to the string equations (upto
field redefinitions), instead of relying on indirect arguments of gauge
invariance and general covariance. 

If indeed these equations are equivalent to string theory equations, then this
is a step towards a background independent formulation of string theory \cite{W}.  The
loop variable approach is not tied down to any particular background. It is also 
important to find out whether there is an action principle here analogous
to that discussed in \cite{W}. Some aspects of this action 
have been discussed in \cite{Poly}.  
   
It would also be interesting to compare these results with other
results on higher spin interactions in curved space\cite{AD,BBB,BBD,BGKP,FV,V}.

\end{document}